\documentclass[preprint,eqsecnum,preprintnumbers,nofootinbib,byrevtex,prd,aps,showpacs,showkeys,groupedaddress,floatfix]{revtex4}
\usepackage{bm}
\usepackage{graphics}
\usepackage{graphicx}
\usepackage{epsfig}
\usepackage{amssymb}
\usepackage{amsmath}

\newcommand\nn{\nonumber}
\newcommand\ba{\begin{eqnarray}}
\newcommand\ea{\end{eqnarray}}

\begin{document}

\date{}
\title{Analytical Solutions Of The  Schr\"{o}dinger
Equation  For The Hulth\'en  Potential Within SUSY Quantum
Mechanics}
\author{H.~I.~Ahmadov$^{1}$}\email{E-mail:hikmatahmadov@yahoo.com}
\author{Sh. I.~Jafarzade$^{2}$}\email{E-mail:shahriyarjafarzade@gmail.com}
\author{M.~V.~Qocayeva$^{3}$} \email{E-mail:mefkureqocayeva@yahoo.com}

\affiliation{$^{1}$ Department of Equation of Mathematical
Physics,Faculty of Applied Mathematics and Cybernetics, Baku State
University, Z. Khalilov st. 23, AZ-1148, Baku, Azerbaijan}
\affiliation{$^{2}$ Particle Physics Research Group, Institute of
Radiation Problems, B.Vahabzade str., 9, AZ1143, Baku, Azerbaijan}
\affiliation{$^{3}$ Institute of Physics, Azerbaijan National
Academy of Sciences, H. Javid Avenue,131, AZ-1143, Baku, Azerbaijan}


\begin{abstract}
The analytical solution of the modified radial Schr\"{o}dinger
equation for the Hulth\'en potential is obtained within ordinary
quantum mechanics by applying the Nikiforov-Uvarov method and
supersymmetric quantum mechanics by applying the shape invariance
consept that was introduced by Gendenshtein method by using the
improved approximation scheme to the centrifugal potential for
arbitrary $l$ states. The energy levels are worked out and the
corresponding normalized eigenfunctions are obtained in terms of
orthogonal polynomials for arbitrary $l$ states.
\end{abstract}


\maketitle

\section{\bf Introduction}

 As known, one of the main objectives in theoretical physics since
the early years of quantum mechanics (QM) is to obtain an exact
solution of the Schr\"{o}dinger equation for some  special
physically  important potentials. Since the wave function contains
all necessary information for full description of a quantum system,
an analytical solution of the Schr\"{o}dinger equation is of high
importance in non-relativistic and relativistic quantum
mechanics~\cite{Greiner,Bagrov}.  There are few potentials for which
the Schr\"{o}dinger equation can be solved explicitly for all $n$
and $l$ quantum states.

The Hulth\'en potential is one of the important short-range
potentials in physics. The potential has been used in nuclear and
particle physics, atomic physics, solid-state physics, and its bound
state and scattering properties have been investigated by a variety
of techniques. General wave functions of this potential have been
used in solid-state and atomic physics problems. It should be noted
that, Hulth\'en potential is a special case of Eckart potential
~\cite{Eckart}.

The  Hulth\'en potential is defined by~\cite{Hulten1,Hulten2}
\begin{equation}
V(r)=-\frac{Ze^2\delta e^{-\delta r}}{(1-e^{-\delta r})}
\end{equation}
where $Z$ is a constant and  $\delta$ is the screening parameter,
dimensionless parameters. It should be noted that, the radial
Schr\"{o}dinger equation for the Hulth\'en potential can be solved
analytically for only the states with zero angular momentum
~\cite{Hulten1,Hulten2,Flugge,Lam,Varshni}. For any $l$ states a
number of methods have been employed to evaluate bound-state
energies
numerically~\cite{Lai1,Dutt,Patil,Popov,Roy,Tang,Lai2,Matthys,Laha,Talukdar,Filho}.

At small values of the radial coordinate $r$, the Hulth\'en
potential behaves like a Coulomb potential,  whereas for large
values of $r$ it decreases exponentially so that its influence for
bound state is smaller than, that of Coulomb potential.

Because of these results, in this article we have used a method
within the frame of supersymmetric quantum mechanics (SUSYQM) using
an effective Hulth\'en potential for any $l\neq 0$ angular momentum
states, which can be solved analytically. In Ref.~\cite{Gonul}
authors used SUSY QM  Hamiltonian hierarchy method for analytically
solving radial Schr\"{o}dinger equation for the Hulth\'en potential
for any $l$ states.

In contrast to the Hulth\'en potential, the Coulomb potential  is
analytically solvable for any $l$ angular momentum. Take into
account of this point will be very interesting and important solving
Schr\"{o}dinger equation for the Hulth\'en potential for any $l$
states within ordinary and  SUSY QM and also to compare and analyze.

In this study, we obtain the energy eigenvalues and corresponding
eigenfunctions for arbitrary $l$ states by solving the
Schr\"{o}dinger equation for the Hulth\'en potential using
Nikiforov-Uvarov (NU) method ~\cite{Nikiforov} and the shape
invariance concept that was introduced by Gendenshtein
~\cite{Gendenshtein} .

It is known that, using for this potential the Schr\"{o}dinger
equation can be solved exactly for s-wave ($l = 0$) ~\cite{Flugge}.

Unfortunately, for an arbitrary $l$-states ($l\neq 0$), the
Schr\"{o}dinger equation does not get an exact solution. But many
papers show the power and simplicity of NU method in solving central
and noncentral potentials~\cite{Badalov1,Badalov2,Badalov3,
Ahmadov1,Ahmadov2} for arbitrary $l$ states. This method is based on
solving the second-order linear differential equation by reducing to
a generalized equation of hypergeometric-type which is a
second-order  type  homogeneous differential equation with
polynomials coefficients of degree not exceeding the corresponding
order of differentiation.

In this study, we obtain the energy eigenvalues and corresponding
eigenfunctions for arbitrary $l$ states by solving the radial
Schr\"{o}dinger equation for the Hulth\'en potential within ordinary
and SUSY QM.

It should be noted that the same problem have been studied within
SUSY QM in Ref.~\cite{Feizi} as well, but our results disagree with
the result  obtained.

The structure  of this work is as follows.  Bound-state Solution of
the radial Schr\"{o}dinger equation for Hulth\'en potential by NU
method within ordinary quantum mechanics is provided in Section
\ref{ir}. The Solution of Schr\"{o}dinger equation for Hulth\'en
potential within SUSY QM \ref{ar} and the numerical results for
energy levels and the corresponding normalized eigenfunctions are
presented in Section \ref{br}. Finally, some concluding remarks are
stated in Section \ref{dr}.

\section{\bf Bound state Solution of the Radial Schr\"{o}dinger
equation for Hulth\'en potential  within ordinary quantum
mechanics.}\label{ir}

The Schr\"{o}dinger equation in spherical coordinates is given as

\ba
\bigtriangledown ^2 \psi+  \frac{2\mu}{\hbar^2}[E-V(r)]\psi=0.
\ea
Considering this equation, the total wave function is written as
\ba
\psi (r,\theta ,\phi )=R(r)Y_{l,m}(\theta,\phi ),
\ea

Thus, for radial Schr\"{o}dinger equation with Hulth\'en potential
is
%
\ba R''(r)+\frac{2}{r}R'(r)+\frac{2\mu }{\hbar
^2}\left[E+Ze^{2}\delta\frac{e^{-\delta r}}{1-e^{-\delta
r}}-\frac{l(l+1)\hbar^2}{2\mu r^2} \right]R(r)=0, \ea

respectively.

The effective Hulth\'en potential is

\ba V_{eff}(r)=V_{H}+V_{l}= -Ze^{2}\delta\frac{e^{-\delta
r}}{1-e^{-\delta r}}+\frac{l(l+1)\hbar^2}{2\mu r^2}, \ea

As we know, Eq.(2.3) is the radial Schr\"{o}dinger equation for
Hulth\'en potential. In order to solve Eq.(2.3) for $l\neq0,$ we
must make an approximation for the centrifugal term. When $\delta r
< < 1$, we use an improved approximation scheme~\cite{Jia} to deal
with the centrifugal term,
\begin{equation}
\left[C_0+\frac{{e^{ - \delta r} }}{{(1 - e^{ - \delta r} )^2 }}
\right] \approx \frac{{1 }}{{\delta^2 r^2 }} + \left( {C_0 -
\frac{1}{{12}}} \right) + O({\delta^2 r^2}),\,\,C_0 =
\frac{1}{{12}},\frac{1}{{r^2 }} \approx \delta^2\left[ {C_o +
\frac{{e^{ - \delta r} }}{{(1 - e^{ - \delta r} )^2 }}} \right],
\end{equation}

Now, the effective potential becomes

\ba\widetilde{V}_{eff}(r)= -Ze^{2}\delta\frac{e^{-\delta
r}}{1-e^{-\delta r}}+\frac{l(l+1)\hbar^2}{2\mu}\left(C_o +
\frac{e^{-\delta r}}{(1 - e^{-\delta r})^2}\right)\ea

where the parameter $C_0=\frac{1}{12}$ (Ref. ~\cite{Jia}) is a
dimensionless constant. However, when $C_0=0$, the approximation
scheme becomes the convectional approximation scheme suggested by
Greene and Aldrich~\cite{Greene}. It should be noted that this
approximation, is only valid for small $\delta r$ and it breaks down
in the high screening region. After using this approximation radial
Schr\"{o}dinger equation is solvable analytically.

We assume $R(r)=\frac{1}{r}\chi(r)$ in Eq.(2.3) and the radial
Schr\"{o}dinger equation becomes
%
\ba \chi''(r) + \frac{2\mu }{\hbar ^2}\left[\varepsilon
+Ze^{2}\delta\frac{e^{-\delta r}}{1-e^{-\delta r}}-\frac{\hbar ^2
l(l+1)}{2\mu}(C_o + \frac{e^{-\delta r}}{(1 - e^{-\delta r})^2
})\right]\chi (r) = 0. \ea
%
In order to transform Eq.(2.7), the equation of the generalized
hypergeometric-type which is in the form ~\cite{Nikiforov}
%
\ba
\chi''(s)+\frac{\tilde{\tau}}{\sigma } \chi'(s)+\frac{\tilde{\sigma}}{\sigma ^2} \chi(s)=0,
\ea

we use  the following ansatz in order to make the differential
equation more compact,
\ba -\varepsilon ^2=\frac{2\mu }{\hbar ^2\delta^2}E,\,\,\,\, E<0,\,\,\,\,
\alpha^2=\frac{2\mu Ze^2}{\hbar^2\delta},\,\,\,\, s = e^{-\delta r}. \ea

Hence, we obtain

\ba \chi''(s)+\frac {\chi'(s)}{s}+ 
\frac{1}{s^2(1-s)^2}\biggr[-\varepsilon^2(1-s)^2-l(l+1)(C_{0}(1-s)^2+s)+\alpha^{2}s(1-s)\biggr]\chi(s)=0.   \nn \\
\ea

Now, we can successfully apply NU method for defining eigenvalues of
energy. By comparing Eq.(2.10) with Eq.(2.8), we can define the
following:

$\tilde{\tau} (s) = 1 - s,\sigma (s) = s(1 - s) $,
%
\ba
\tilde{ \sigma} (s) = - \epsilon
^2(1-s^2)-l(l+1)(C_{0}(1-s)^2+s))+\alpha^{2}s(1-s).
\ea

We change $\lambda=l(l+1)$, then we obtain:
%
\ba \tilde{ \sigma} (s) = - \epsilon ^2(1-s^2)-\lambda(C_{0}(1-s)^2+
s))+\alpha^{2}s(1-s). \ea

If we take the following factorization,
%
\ba
\chi (s)=\phi (s)y(s),
\ea
for the appropriate function $\phi (s)$, Eq.(2.10) takes the form of
the well-known hypergeometric-type equation. The appropriate $\phi
(s)$ function must satisfy the following condition:
%
\ba
\frac { \phi ^{'} (s)}{\phi (s)}=\frac {\pi (s)}{\sigma (s)},
\ea
where function $\pi (s)$ is defined as
%
\ba
\pi(s)= \frac{{ \sigma' -\tilde{\tau }}}{2} \pm \sqrt {  (\frac{{ \sigma' -\tilde{\tau} }}{2} )^2 -\tilde{\sigma} +k\sigma }.
\ea
Finally, the equation, where $y(s)$ is one of its solutions, takes
the form known as hypergeometric-type,
%
\ba
\sigma (s) y^{''} (s) + \tau (s) y^{'} (s) +\bar{\lambda} y(s)=0,
\ea
where
\ba
\bar{\lambda} =k+\pi ^{'}
\ea
and
\begin{equation}
  \tau (s)=\tilde{\tau} (s) +2\pi (s).
\end{equation}

For our problem, the $\pi (s)$ function is written as
\ba
\pi(s)= \frac{{ - s}}{2} \pm \sqrt {s^2 [a - k] - s[b -k] + c},
\ea

where the values of the parameters are
\ba
a = \frac{1}{4} +\epsilon ^2 + \lambda C_0 + {\alpha}^2, \nn \\
b = 2\epsilon ^2 + 2\lambda C_0 + {\alpha}^2 -\lambda,  \nn \\
c = \varepsilon ^2  + \lambda C_0. \nn \ea

The constant parameter $k$ can be found under the condition that the
discriminant of the expression under the square root is equal to
zero. Hence, we obtain
\ba
k_{1,2}  = (b - 2c) \pm 2\sqrt {c^2  + c(a - b)}.
\ea

Now, we can find four possible functions for $\pi(s)$:

\ba
\pi (s) = \frac{{ - s}}{2} \pm \left\{ \begin{array}{l}
(\sqrt c  - \sqrt {c + a - b} )s - \sqrt c \,\,\,for\,\,\, k = (b - 2c) + 2\sqrt {c^2  + c(a - b)} , \\
(\sqrt c  + \sqrt {c + a - b} )s - \sqrt c \,\,\, for\,\,\, k = (b - 2c) - 2\sqrt {c^2  + c(a - b)} . \\
\end{array} \right.
\ea

According to NU method, from the four possible forms of the
polynomial $\pi(s)$, we select the one for which the function $\tau
(s)$  has the negative derivative. Therefore, the appropriate
function $\pi(s)$ and $\tau(s)$ are

\ba
\pi^{'}(s)=-\frac{1}{2}-\left[\sqrt{c}+\sqrt{c+a-b}\right],
\ea
\ba
\pi(s)=\sqrt{c}-s\left[\frac{1}{2}+\sqrt{c}+\sqrt{c+a-b}\right],
\ea
\ba \tau (s) = 1 +2\sqrt c -2s \left[ 1+\sqrt c+\sqrt{c+a-b}\right]
, \ea

for
\ba
k  = (b - 2c) -2\sqrt {c^2  + c(a - b)}.
\ea

Also by Eq.(2.17), we can define the constant $ \bar{\lambda} $ as
\ba
\bar{ \lambda}=b-2c- 2\sqrt {c^2  + c(a - b)}  - \left[\frac{1}{2} + {\sqrt{c} + \sqrt {c + a - b} }\right].
\ea

Given a nonnegative integer $n$, the hypergeometric-type equation
has a unique polynomial solution of degree $n$ if and only if

\ba
\bar{\lambda}=\bar{\lambda}_n=-n\tau'-\frac{n(n-1)}{2}\sigma '', \,\,\,(n=0,1,2...)
\ea

and $\bar{\lambda}_m\neq\bar{\lambda}_n $ for
~$m=0,1,2,...,n-1$~\cite{Area}, then it follows that
\ba
\bar{\lambda} _{n_{r}}  = b-2c- 2\sqrt {c^2  + c(a - b)}
-\left[\frac{1}{2} + {\sqrt{c} + \sqrt {c + a - b} }\right] = \nn \\
2n_r\left[ {1 +\left( {\sqrt c  + \sqrt {c + a - b} } \right)}
\right] + n_r(n_r - 1).
\ea

We can solve Eq.(2.28) explicitly for $c$ and by using the relation
 $c=\varepsilon ^2  + \lambda C_0$, which brings

\ba \varepsilon^{2}  = \left[
\frac{\lambda+1/2+(l+\frac{1}{2})(2n+1)+2n+n^2-n-\alpha^2}{2(l+\frac{1}{2})+2n+1}
\right]^2  -\lambda
  C_0,
\ea

Finally, we can found for $\varepsilon^{2}$

\ba \varepsilon^{2}  = \left[
\frac{l+n+1}{2}-\frac{\alpha^2}{2(l+n+1)} \right]^2 -l(l+1)C_0. \ea

We substitute $\varepsilon ^2$ into Eq.(2.9) with $\lambda =l(l+1)$,
which identifies

\ba E_{nl} = \frac{-h^2 }{2\mu
}\left[\frac{(l+n+1)}{2}\delta-\frac{\frac{\mu Z}{\hbar
^2}e^2}{l+n+1} \right]^2 +\frac{\hbar ^2 \delta^2}{2\mu}l(l+1) C_0.
\ea

If we take $C_0=0$ in the Eq.(2.31), then we obtain result
~\cite{Ikhdair}.

Now, using NU method we can obtain the radial eigenfunctions. After
substituting   $\pi(s)$ and $\sigma(s) $ into Eq.(2.14) and solving
first-order differential equation, it is easy to obtain

\ba \phi (s)=s^{\sqrt c}(1-s)^{K},
 \ea
where $K=\frac{1}{2}+ \sqrt {c+a-b}=l+1$

Furthermore, the other part of the wave function y(s) is the
hypergeometric-type function whose polynomial solutions are given by Rodrigues relation
\ba
y_{n}(s) = \frac {B_{n}}{\rho (s)} \frac{{d^{n} }}{{ds^{n}
}}\left[ \sigma ^{n}(s)\rho (s) \right],
\ea
where $B_n$ is a normalizing constant and $\rho (s)$ is the weight
function which is the solution of the Pearson differential equation.
The Pearson differential equation and $\rho(s)$ for our problem is
given as

\ba
(\sigma \rho )^{'} =\tau \rho ,
\ea

\ba \rho(s) =(1 -s)^{2k - 1} s^{2\sqrt c }, \ea
respectively.

Substituting Eq.(2.35) in Eq.(2.33) we get

\ba
y_{n_{r}}(s) = B_{n_{r}}(1 - s)^{1 - 2K} s^{2\sqrt c }
\frac{{d^{n_{r}} }}{{ds^{n_{r}} }}\left[ {s^{2\sqrt c  + n_{r}} (1 -
s)^{2K - 1 + n_{r}} } \right].
\ea
Then, by using the following definition of the Jacobi  polynomials
~\cite{Abramowitz}:

\ba
P_n^{(a,b)} (s) = \frac{( - 1)^n }{n!2^n (1 - s)^a (1 + s)^b}\frac{d^n }{ds^n }\left[ {(1 - s)^{a + n} (1 + s)^{b + n} }
\right],
\ea

we can write

\ba
P_n^{(a,b)} (1-2s) = \frac{C_n}{ s^a (1 - s)^b}\frac{d^n }{ds^n }\left[s^{a+n}(1-s)^{b+n}\right]
\ea

and

\ba
\frac{d^n }{ds^n }\left[s^{a+n}(1-s)^{b+n}\right]=C_n s^a (1 - s)^b P_n^{(a,b)} (1-2s).
\ea

If we use the last equality in Eq.(2.36), we can write

\ba
y_{n_{r}}(s) = C_{n_{r}} P_{n_{r}}^{(2\sqrt{c},2K-1)} (1-2s).
\ea

Substituting $\phi (s)$ and $y_{n_{r}}(s)$ into Eq.(2.13), we obtain

\ba
\chi _{n_{r}}(s)=C_{n_{r}}s^{\sqrt c}(1-s)^K P_{n_{r}}^{(2\sqrt{c},2K-1)} (1-2s).
\ea

Using the following definition of the Jacobi
polynomials~\cite{Abramowitz}:

\ba
P_n^{(a,b)} (s) = \frac{{\Gamma (n + a + 1)}}{{n!\Gamma (a +
1)}}\mathop F\limits_{21} \left( { - n,a + b + n + 1,1 + a;\frac{{1
- s}}{2}} \right),
\ea

we  are able to write Eq.(2.41) in terms of hypergeometric
polynomials as

\ba
\chi_{n_{r}} (s)=C_{n_{r}}s^{\sqrt c}(1-s)^{K}\frac{\Gamma (n_{r}+2\sqrt c+1)}{n_{r}!\Gamma (2 \sqrt c+1)} \mathop F\limits_{21} \left( { - n_{r},2 \sqrt c +2K+n_{r},1 +2 \sqrt
c;s}\right).
\ea

The normalization constant $C_{n_{r}}$ can be found from
normalization condition

\ba
\int\limits_0^\infty |R(r)|^2r^2dr=\int\limits_0^\infty |\chi (r)|^2 dr=b\int\limits_0^1\frac{1}{s}|\chi (s)|^2 ds=1,
\ea

by using the following integral formula~\cite{Agboola}:
\ba
\int\limits_0^1 {(1 - s)^{2(\delta  + 1)} s^{2\lambda  - 1} } \left\{ {\mathop F\limits_{21} ( - n_{r},2(\delta  + \lambda  + 1) + n_{r},2\lambda  + 1;s)} \right\}^2 dz = \nn \\
\frac{{(n_{r} + \delta  + 1)n_{r}!\Gamma (n_{r} + 2\delta  + 2)\Gamma (2\lambda )\Gamma (2\lambda  + 1)}}{{(n_{r} + \delta  + \lambda  + 1)\Gamma (n_{r} + 2\lambda  + 1)\Gamma (2(\delta  + \lambda  + 1) + n_{r})}},
\ea
for $ \delta  > \frac{{ - 3}}{2}$\,\,\, and\,\,\, $\lambda >0 $.
After simple calculations, we obtain normalization constant as
\ba
C_{n_r}=\sqrt{\frac{n_{r}!2\sqrt c(n_{r}+K+\sqrt c)\Gamma (2(K+\sqrt c)+n_{r})}{b(n_{r}+K)\Gamma (n_{r}+2\sqrt c+1)\Gamma (n_{r}+2K)} }.
\ea

\section{\bf The Solution of Schr\"{o}dinger equation for Hulth\'en potential within SUSY Quantum Mechanics}\label{ar}

In the Supersymmetric QM, it is necessary to define nilpotent
operators, namely $Q$ and  $Q_{+}$, satisfying the algebra

\ba \label{eq:MassMatrixC}
Q=\left(\begin{array}{cc}0& 0\\
A^{-}&0\end{array}\right), \ea

\ba \label{eq:MassMatrixC}
Q_{+}=\left(\begin{array}{cc}0& A^+\\
0&0\end{array}\right), \ea

where $A^{+}$ and $A^{-}$ are bosonic operators.

The Hamiltonian, $H$ in terms of these operators is given by
\ba
\label{eq:MassMatrix1}
H=\left(\begin{array}{cc}A^{+}A^{-}& 0\\
0&A^{+}A^{-}\end{array}\right)=\left(\begin{array}{cc}H_{+}& 0\\
0&H_{-}\end{array}\right).
\ea

Supersymmetric algebra allows us to write Hamiltonians
as~\cite{Cooper1,Cooper2}
\ba
H_{\pm}=-\frac{{\hbar}^{2}}{2\mu}\frac{d^{2}}{dx^{2}}+V_{\pm}(x),
\ea
where the SUSY partner potentials $V_{\pm}$ in terms of the
superpotential $W(x)$
\ba
 V_{\pm}=W^2\pm\frac{\hbar}{2\mu}\frac{dW}{dx}
\ea

The superpotential has a definition
\ba
W(x)=-\frac{\hbar}{\sqrt{2\mu}}\left(\frac{d\ln\psi_{0}^{(0)}(x)}{dx}\right),
\ea

where $\psi_{0}^{(0)}(x)$ denotes the ground-state wave function
that satisfies the relation
\ba
\psi_{0}^{(0)}(x)=N_{0}exp\left(-\frac{\sqrt{2\mu}}{\hbar}\int^{x}W(x^{'})dx^{'}\right).
\ea

The Hamiltonian $H_{\pm}$ can also be written in terms of the
bosonic operators $A^{+}$ and $A^{-}$
\ba
H_{\pm}=A^{\mp}A^{\pm}
\ea
where
\ba A^{\pm}= -\frac{\hbar}{\sqrt{2\mu}}\frac{d}{dx}+W(x) \ea
It is a remarkable result that the energy eigenvalues of $H_{-}$ and
$H_{+}$ are identical except for the ground-state. In the case of
unbroken supersymmetry, the ground-state energy of the Hamiltonian
$H_{-}$ is zero $E_{0}^0=0$ ~\cite{Cooper1,Cooper2}. In the
factorization of the Hamiltonian, Eqs.(3.4), (3.8) and (3.9) are
used, respectively. Hence, we obtain
\ba H_{1}=-\frac{{\hbar}^{2}}{2\mu}\frac{d^{2}}{dx^{2}}+V_{1}(x)=
A_{1}^{+}A_{1}^{-}+E_{0}^{(1)} \ea
Thus, comparing each side of Eq.(3.10), term by term, we receive the
Riccati equation for the superpotential $W_{1}$
\ba
W_{1}^{2}(x)-W_{1}^{'}(x)=\frac{2\mu}{{\hbar}^{2}}(V_{1}(x)-E_{0}^{(1)}).
\ea
Let us now construct the SUSY partner Hamiltonian $H_{2}$ as
\ba H_{2}=-\frac{{\hbar}^{2}}{2\mu}\frac{d^{2}}{dx^{2}}+V_{2}(x)=
A_{2}^{+}A_{2}^{-}+E_{0}^{(2)} \ea

and Riccati equation takes the form
\ba
W_{2}^{2}(x)-W_{2}^{'}(x)=\frac{2\mu}{{\hbar}^{2}}(V_{2}(x)-E_{0}^{(2)}).
\ea

Similarly, one can write, in general, the Riccati equation and
Hamiltonians by iteration as
\ba
W_{n}^{2}(x)-W_{n}^{'}(x)=\frac{2\mu}{{\hbar}^{2}}(V_{n}(x)-E_{0}^{(2)})=A_{n}^{\pm}A_{\mp}^{-}+E_{0}^{(n)}.
\ea
and
\ba
H_{n}=-\frac{{\hbar}^{2}}{2m}\frac{d^{2}}{dx^{2}}+V_{n}(x)=
A_{n}^{+}A_{n}^{-}+E_{0}^{(n)},   n=1,2,3......
\ea

Because of the SUSY unbroken case, the partner Hamiltonians satisfy
the following expressions~\cite{Cooper1,Cooper2}
\ba E_{0}^{(n+1)}= E_{1}^{(n)}, n=1,2,3......;E_{0}^{(0)}=0 \ea

In SUSY QM, the ground-state eigenfunction $\psi_{0}(x)$ can be
written as Eq.(3.7). Through the superalgebra, we make following
ansatz for the superpotential:

\ba W_{1}(r)=-\frac{\hbar}{\sqrt 2 \mu}\left(A+\frac{Be^{-\delta
r}}{1-e^{-\delta r}}\right) \ea

and having inserted this expression  into Eq.(3.11), we obtain

\ba && W_{1}^{2}(r)-\frac{\hbar}{\sqrt {2\mu}}W^{'}_{1}(r)  =
\frac{{\hbar}^{2}}{2m}\left(A^2+\frac{2ABe^{-\delta r}}{1-e^{-\delta
r}}+\frac{B^2 e^{-2\delta r}}{(1-e^{-\delta r})^2}-\frac{\delta B
e^{-\delta r}}{(1-e^{-\delta r})^2}\right) \ea
If take into account Eqs.(2.7) and (3.18), then we obtain:

\ba \frac{{\hbar}^{2}}{2\mu}\left(A^2+\frac{2ABe^{-\delta
r}}{1-e^{-\delta r}}+\frac{B^2 e^{-2\delta r}}{(1-e^{-\delta
r})^2}-\frac{\delta B e^{-\delta r}}{(1-e^{-\delta r})^2}\right)= \nn \\
\left[-\varepsilon -\frac{Ze^{2}\delta e^{-\delta r}}{1-e^{-\delta
r}} +\frac{\hbar ^2 l(l+1)}{2\mu}\delta^{2}C_o + \frac{\hbar ^2
l(l+1)}{2\mu}\frac{\delta^{2}e^{-\delta r}}{(1-e^{-\delta
r})^2}\right]. \ea

After small manipulations, we obtain

\ba A^2+\frac{2ABe^{-\delta r}}{1-e^{-\delta r}}+\frac{B^2
e^{-2\delta r}}{(1-e^{-\delta
r})^2}-\frac{\delta B e^{-\delta r}}{(1-e^{-\delta r})^2}= \nn \\
\frac{2\mu}{{\hbar}^{2}}\left[-E -\frac{Ze^{2}\delta e^{-\delta
r}}{1-e^{-\delta r}} +\frac{\hbar ^2 l(l+1)}{2\mu}\delta^{2}C_o +
\frac{\hbar ^2 l(l+1)}{2\mu}\frac{\delta^{2}e^{-\delta
r}}{(1-e^{-\delta r})^2}\right] \ea

where it satisfies the associated Riccati equation, so we can obtain
the following identity.

With comparison  of the each side of the Eq.(3.20), we obtain

\ba A^2=-\frac{2\mu E}{{\hbar}^{2}}+\delta^{2}
C_{0}l(l+1)=\varepsilon^2\delta^2+\delta^2C_{0}l(l+1), \ea \ba
2AB-\delta B=\delta^2
l(l+1)-\frac{2\mu}{{\hbar}^{2}}Ze^2\delta=\delta^2l(l+1)-\delta^2\alpha^2,
\ea \ba B^2-\delta B = \delta^2l(l+1). \ea

After inserting Eq.(3.17) into (3.7), the eigenfunction for
ground-state in terms of $r$ will be obtained as
\begin{equation}
\chi(r)= N_{0}e^{A r}(1-e^{-\delta r})^{\frac{B}{\delta}}
\end{equation}

Considering extremity conditions to wave functions, we obtain $B>0$
and  $A<0$.

Solving Eq.(3.23) yields

\begin{equation}
B=\frac{\delta\pm\sqrt{\delta^2+4\delta^2
l(l+1)}}{2}=\frac{\delta\pm\delta\sqrt{(2l+1)^2}}{2}=\delta\pm
\delta(2l+1),
\end{equation}

and considering $B>0$ from Eqs.(3.22) and (3.23), we find

\begin{equation}
2AB-B^2=-\delta^2 \alpha^2,
\end{equation}

or

\begin{equation}
A=\frac{B}{2}-\frac{\delta^2 \alpha^2}{2B},
\end{equation}

From Eqs.(2.9) and (3.21), we find

\begin{equation}
-\frac{2\mu
E_0}{{\hbar}^{2}\delta^2}=\frac{1}{\delta^2}\left(\frac{B}{2}-\frac{\delta^2
\alpha^2}{2B}\right)^2- C_{0}l(l+1).
\end{equation}

Finally, for energy eigenvalue, we obtain

\begin{equation}
E_0=\frac{{\hbar}^{2}l(l+1)C_0\delta^2}{2\mu}-
\frac{{\hbar}^2}{2\mu}\biggl(\frac{l+1}{2}\delta-\frac{\alpha^2
\delta}{2(l+1)}\biggr)^2,
\end{equation}

Using Eq.(3.5), we can find SUSY partner potentials $V_{+}(r)$ and
$V_{-}(r)$ in the form

\ba
V_{+}(r) &=& W^2 (r) + \frac{\hbar }{{\sqrt {2\mu } }}W'(r) = \nn \\
&&\frac{{\hbar ^2 }}{{2\mu }}\left[ A^2  + \frac{(2AB +\delta
B)e^{-\delta r}}{1 - e^{ - \delta r}} + \frac{(B^2+\delta
B)e^{-2\delta r}}{(1 - e^{ - \delta r} )^2} \right]\ea

\ba
V_{-}(r) &=& W^2 (r) + \frac{\hbar }{{\sqrt {2\mu } }}W'(r) = \nn \\
&&\frac{{\hbar ^2 }}{{2\mu }}\left[ A^2  + \frac{(2AB -\delta
B)e^{-\delta r}}{1 - e^{ - \delta r}} + \frac{(B^2-\delta
B)e^{-2\delta r}}{(1 - e^{ - \delta r} )^2} \right]\ea

The shape invariance concept that was introduced by Gendenshtein
is~\cite{Gendenshtein}
\ba R(B_1 ) = V_{+}(B,r) - V_{-}(B_1 ,r) = \frac{{\hbar ^2 }}{{2\mu
}}\left[A^2 -A_{1}^2\right] =  \nn \\
\frac{{\hbar ^2 }}{{2\mu }}\left[
\left(\frac{B}{2}-\frac{\delta ^2 \alpha^2}{2B}\right)^2 -
\left(\frac{B + \delta}{2}-\frac{\delta ^2 \alpha^2}{2(B+\delta)}
\right)^2\right].
 \ea
If we now consider a mapping of the form
\ba
B \to B_{1} = B + \delta, \nn \\
B_{n}  = B + n\delta, \ea
so, we have
\ba
R(B_i ) = V_ +  [B + (i - 1)\delta ,r] - V_ -  [B + i\delta ,r] = \nn \\
-\frac{{\hbar ^2 }}{{2\mu }}\left[
\left(\frac{B+i\delta}{2}-\frac{\delta ^2
\alpha^2}{2(B+i\delta)}\right)^2 - \left(\frac{B +
(i-1)\delta}{2}-\frac{\delta ^2 \alpha^2}{2(B+(i-1)\delta)}
\right)^2\right].
 \ea

where the reminder $R(B_1)$ is independent of $r$. Thus, we have
\ba
E_{nl} &=& E_0  + \sum\limits_{i = 0}^n R(B_i ) = \nn \\
&&\frac{{\hbar}^2l(l+1)}{{2\mu }} \delta^2 C_0 - \frac{{\hbar^2
}}{{2\mu}}\left(\frac{B}{2} - \frac{\delta^2 \alpha^2}{2B}
\right)^2- \frac{{\hbar^2 }}{{2\mu}}\biggl[\left(\frac{B+\delta}{2}
\frac{\delta^2 \alpha^2}{2(B+\delta)} \right)^2-  \nn \\
&&\left(\frac{B}{2} - \frac{\delta^2 \alpha^2}{2B} \right)^2 +
\left(\frac{B+2\delta}{2} -
\frac{\delta^2 \alpha^2}{2(B+2\delta)}\right)^2+.....+ \nn \\
&&\left(\frac{B+(n-1)\delta}{2} -
\frac{\delta^2\alpha^2}{2(B+(n-1)\delta)}\right)^2 -
\left(\frac{B+(n-2)\delta}{2} - \frac{\delta^2
\alpha^2}{2(B+(n-2)\delta)}\right)^2-  \nn \\
&&\left(\frac{B+(n-2)\delta}{2} - \frac{\delta^2
\alpha^2}{2(B+(n-2)\delta)}\right)^2 +
\left(\frac{B+n\delta}{2} -
\frac{\delta^2\alpha^2}{2(B+n\delta)}\right)^2- \nn \\
&&\left(\frac{B+(n-1)\delta}{2} -
\frac{\delta^2\alpha^2}{2(B+(n-1)\delta)}\right)^2\biggr] =  \nn \\
&&\frac{{\hbar}^2l(l+1)}{{2\mu}}\delta^2 C_0-
\frac{{\hbar^2}}{{2\mu}} \left(\frac{B+n\delta}{2} -
\frac{\delta^2\alpha^2}{2(B+n\delta)} \right)^2, \ea

and we obtain

\ba E_{nl} &=& \frac{{\hbar}^2l(l+1)}{{2\mu}}\delta^2 C_0-
\frac{{\hbar^2}}{{2\mu}} \left(\frac{B+n\delta}{2} -
\frac{\delta^2\alpha^2}{2(B+n\delta)} \right)^2, \ea

Finally, for energy eigenvalues we found

\ba E_{nl}= \frac{{\hbar}^2l(l+1)}{{2\mu}}\delta^2 C_0-
\frac{{\hbar^2}}{{2\mu}} \left[{\frac{n+l+1}{2}}{\delta} -
\frac{\frac{\mu Ze^2}{{\hbar}^2}}{(l+n+1)}\right]^2, \ea

\section{\bf Numerical Results and Discussion}\label{br}

Solution of the modified radial Schr\"{o}dinger equation for the
Hulth\'en potential are obtained within ordinary quantum mechanics
by applying the Nikiforov-Uvarov method and within SUSY QM by
applying the shape invariance concept that was introduced by
Gendenshtein method in which we have used the improved approximation
scheme to the centrifugal potential for arbitrary $l$ states. Both
ordinary and SUSY quantum mechanical energy eigenvalues and
corresponding eigenfunctions  have obtained for arbitrary $l$
quantum numbers. In the Table I, we present numerical results for
the energy eigenvalues of the Hulth\'en potential as a function of
screening parameter for various state in atomic units is obtained by
within ordinary (obtained by NU method) and SUSY QM(shape invariance
method) methods.

For comparison, in the Table II shows that energy eigenvalues of the
Hulth\'en potential as a function of screening parameter for various
state in atomic units which are obtained of the asymptotic iteration
method ~\cite{Boztosun}, the SUSY ~\cite{Gonul}, numerical
integration ~\cite{Varshni} and the variational method
~\cite{Varshni}. As it can be seen  from the results presented in
these tables, the numerical results obtained of the  analytically
solution are in good agreement with results of the other methods for
the small $\delta$ values, but in the large screening region, the
agreement is poor. Analysis our calculation is shows that the main
reason is simply that when the $\delta r$ increases in the large
screening region, the agreement between $V_{eff}(r)$ and
$\widetilde{V}_{eff}(r)$ potential decreases. However, this problem
could be solved by making a better approach of the centrifugal term.

It should be noted, that Eqs.(2.31)and (3.37) in cases $C_0=0$ and
$l\neq 0$ is exactly the same result obtained by other works
~\cite{Gonul,Ikhdair}, also Eqs.(2.31) and (3.37) in cases $C_0=0$
and $l = 0$ is exactly the same result in ~\cite{Flugge}.

It is shown, that energy eigenvalues and corresponding
eigenfunctions are identical for both ordinary and SUSY QM.

\section{\bf Conclusion}\label{dr}

It is well know that the Hulth\'en potential is one of the important
exponential potential, and it has been a subject of interest in many
fields of physics and chemistry. The main results of this paper are
the explicit and closed form expressions for the energy eigenvalues
and the normalized wave functions. The method presented in this
paper is a systematic one and in many cases it is more than the
other ones.

Analytical solution of the modified radial Schr\"{o}dinger equation
for the Hulth\'en potential are obtained within ordinary quantum
mechanics by applying the Nikiforov-Uvarov method and within SUSY QM
by applying the shape invariance concept that was introduced by
Gendenshtein method in which we used the improved approximation
scheme to the centrifugal potential for arbitrary $l$ states. The
energy eigenvalues and corresponding eigenfunctions are obtained for
arbitrary $l$ quantum numbers. It is shown that energy eigenvalues
and corresponding eigenfunctions are the same for both ordinary and
SUSY QM.

Consequently, studying of analytical solution of the modified
Schr\"{o}dinger equation for the Hulth\'en potential within
framework ordinary and SUSY QM could provide valuable information on
the QM dynamics at atomic and molecules physics and opens new
window.

We can conclude that our results are not only interesting for pure
theoretical physicist but also for experimental physicist because of
the exact and more general the results.

\section*{Acknowledgments}

The authors  thanks to  Dr. A.I.Ahmadov and Dr. V.H.Badalov for
fruitful discussions and useful comments.

 \newpage

\begin{table}
\begin{tabular}{|c|c|c|c|c|c|c|c|}\hline
 $state $ & $\delta $&Present work, & Present work, & Present work, & Present work, \\
  &&NU $C_0=0$ &NU $C_0\neq0$& SUSY  $C_0=0$ & SUSY  $C_0\neq0$ \\ \hline

  2p&0.025&-0.1128125 &-0.1127604&-0.1128125 &-0.1127604 \\ \hline
  &0.050 &-0.1012500&-0.10104166 &-0.1012500&-0.10104166  \\ \hline
  &0.075&-0.0903125 &-0.08984375 &-0.0903125 &-0.08984375 \\ \hline
  &0.10&-0.080000 &-0.07916666 &-0.080000 &-0.07916666   \\ \hline
  &0.150 &-0.0612500 &-0.059375 &-0.0612500 &-0.059375  \\ \hline
  &0.200 &-0.45000&-0.0416666&-0.45000&-0.0416666  \\ \hline
  &0.250 &-0.0312500 &-0.02604166 &-0.0312500 &-0.02604166  \\ \hline
  &0.300 &-0.02000 &-0.012500 &-0.02000 &-0.012500 \\ \hline
  &0.350 &-0.01125 &-0.00104166 &-0.01125 &-0.00104166  \\ \hline
  3p&0.025 &-0.04375868 &-0.04370659 &-0.04375868 &-0.04370659   \\
  &0.050 &-0.03336805 &-0.03315972 &-0.03336805 &-0.03315972  \\ \hline
  &0.075 &-0.02438737&-0.0239149305 &-0.02438737&-0.0239149305 \\  \hline
  &0.100 &-0.01680555 &-0.015972222 &-0.01680555 &-0.015972222 \\ \hline
  &0.150 &-0.00586805 &-0.003993055 &-0.00586805 &-0.003993055     \\ \hline
  3d&0.025 &-0.04375868 &-0.04370659 &-0.04375868 &-0.04370659  \\  \hline
  &0.050 &-0.03336805&-0.03315972 &-0.03336805 &-0.03315972  \\ \hline
  &0.075 &-0.02438737 &-0.0239149305 &-0.02438737 &-0.0239149305 \\ \hline
  &0.100 &-0.01680555 &-0.015972222 &-0.01680555 &-0.015972222  \\ \hline
  &0.150 &-0.00586805 &-0.003993055 &-0.00586805 &-0.003993055   \\ \hline
  4p&0.025 &-0.02000 &-0.0199478 &-0.02000 &-0.0199478  \\ \hline
    &0.050&-0.01125& -0.011041666&-0.01125& -0.011041666 \\ \hline
    &0.075& -0.00500&-0.00453125 & -0.00500&-0.00453125        \\ \hline
    &0.100&-0.00125&-0.00041666& 0.00125&-0.00041666            \\ \hline
  4d&0.025&-0.02000& -0.0199478& -0.02000& -0.0199478  \\ \hline
    &0.050&-0.01125 &-0.011041666&-0.01125 &-0.011041666\\ \hline
    &0.075&-0.00500 & -0.00453125&-0.00500 & -0.00453125\\ \hline

\end{tabular}
\end{table}

\newpage

\begin{table}
\begin{tabular}{|c|c|c|c|c|c|c|c|}\hline
 $state $ & $\delta $&Present work & Present work & Present work & Present work \\
  &&NU $C_0=0$ & NU $C_0\neq0$& SUSY  $C_0=0$ & SUSY  $C_0\neq0$ \\ \hline

4f&0.025&-0.02000&-0.0199478&-0.02000&-0.0199478\\ \hline
    &0.050&-0.01125 &-0.011041666&-0.01125 &-0.011041666\\ \hline
    &0.075&-0.00500 &-0.00453125&-0.00500 &-0.00453125\\ \hline
  5p&0.025& -0.009453125 &-0.009401&-0.009453125 &-0.009401\\ \hline
    &0.050&-0.0028125 &-0.00260416&-0.0028125 &-0.00260416\\ \hline
5d&0.025& -0.009453125 &-0.009401&-0.009453125 &-0.009401\\ \hline
    &0.050&-0.0028125 &-0.00260416&-0.0028125 &-0.00260416\\ \hline
5f&0.025& -0.009453125 &-0.009401&-0.009453125 &-0.009401\\ \hline
    &0.050&-0.0028125 &-0.00260416&-0.0028125 &-0.00260416\\ \hline
5g&0.025& -0.009453125 &-0.009401&-0.009453125 &-0.009401\\ \hline
    &0.050&-0.0028125 &-0.00260416&-0.0028125 &-0.00260416\\ \hline
  5f&0.025&-0.009453125 &-0.009401&-0.009453125 &-0.009401\\ \hline
    &0.050&-0.0028125 &-0.00260416&-0.0028125 &-0.00260416\\ \hline
  5g&0.025&-0.009453125 &-0.009401&-0.009453125 &-0.009401 \\ \hline
    &0.050&-0.0028125 &-0.00260416&-0.0028125 &-0.00260416\\ \hline
  6p &0.025& -0.00420138&-0.004149305&-0.00420138&-0.004149305\\ \hline
6d &0.025& -0.00420138&-0.004149305&-0.00420138&-0.004149305\\
\hline
6g&0.025& -0.00420138&-0.004149305&-0.00420138&-0.004149305
\\ \hline

\end{tabular}

\caption{Energy eigenvalues of the Hulthe\'n potential as a function
of the screening parameter for
2p, 3p, 3d, 4p, 4d, 4f, 5p, 5d, 5f, 5g, 6p, 6d, 6f and 6g states in atomic units
($\hbar=m=e=1$) and for $Z=1$.} \label{table1}
\end{table}

\clearpage

\begin{table}
\begin{tabular}{|c|c|c|c|c|c|c|c|}\hline
 $state $ & $\delta $&AIM ~\cite{Boztosun} &SUSY~\cite{Gonul} & Numerical~\cite{Varshni} & Variational~\cite{Varshni} \\

  2p&0.025&0.1128125 &0.1127605&0.1127605 &0.1127605 \\ \hline
  &0.050 &0.1012500&0.1010425 &0.1010425&0.1010425  \\ \hline
  &0.075&0.0903125 &0.0898478 &0.0898478 &0.0898478\\ \hline
  &0.10&0.0800000 &0.0791794 &0.0791794 &0.0791794  \\ \hline
  &0.150 &0.0612500 &0.0594415 &0.0594415 &0.0594415  \\ \hline
  &0.200 &0.450000&0.0418854&0.0418860&0.0418860  \\ \hline
  &0.250 &0.0312500 &0.0266060 &0.0266111&0.0266108  \\ \hline
  &0.300 &0.0200000 &0.0137596 &0.0137900 &0.0137878 \\ \hline
  &0.350 &0.0112500 &0.0036146 &0.0037931 &0.0037734  \\ \hline
  3p&0.025 &0.0437590 &0.0437068 &0.0437069 &0.0437069  \\
  &0.050 &0.0333681 &0.0331632 &0.0331645 &0.0331645  \\ \hline
  &0.075 &0.0243837&0.0239331 &0.0239397&0.0239397 \\  \hline
  &0.100 &0.0168056 &0.0160326 &0.0160537 &0.0160537 \\ \hline
  &0.150 &0.00586811 &0.0043599 &0.0044663 &0.0044660     \\ \hline
  3d&0.025 &0.0437587 &0.0436030 &0.0436030 &0.0436030  \\  \hline
  &0.050 &0.0333681&0.0327532 &0.0327532 &0.0327532  \\ \hline
  &0.075 &0.0243837 &0.0230306 &0.0230307 &0.0230307 \\ \hline
  &0.100 &0.0168055 &0.0144832 &0.0144842 &0.0144842 \\ \hline
  &0.150 &0.0058681 &0.0132820 &0.0013966 &0.0013894  \\ \hline
  4p&0.025 &0.0200000 &0.0199480 &0.0199489 &0.0199489 \\ \hline
    &0.050&0.0112500& 0.0110430&0.0110582& 0.0110582 \\ \hline
    &0.075&0.0050000&0.0045385 &0.0046219&0.0046219  \\ \hline
    &0.100&0.0012500&0.0004434& 0.0007550&0.0007532 \\ \hline
  4d&0.025&0.0200000& 0.0198460& 0.0198462& 0.0198462\\ \hline
    &0.050&0.0112500&0.0106609&0.0106674 &0.0106674 \\ \hline
    &0.075&0.0050000 &0.0037916&0.0038345 &0.0038344 \\ \hline

\end{tabular}
\end{table}

\begin{table}
\begin{tabular}{|c|c|c|c|c|c|c|c|}\hline
 $state $ & $\delta $&AIM ~\cite{Boztosun} &SUSY~\cite{Gonul} & Numerical~\cite{Varshni}  & Variational~\cite{Varshni} \\ \hline

 4f&0.025&0.0200000&0.0196911&0.0196911&0.0196911 \\ \hline
    &0.050&0.0112500 &0.0100618&0.0100620 &0.0100620 \\ \hline
    &0.075&0.0050000 &0.0025468&0.0025563 &0.0025557 \\ \hline
  5p&0.025&0.0094531 &0.0094011&0.0094036 & \\ \hline
    &0.050&0.0028125 &0.0026058&0.0026490 & \\ \hline
5d&0.025&0.0094531 &0.0092977&0.0093037 & \\ \hline
    &0.050&0.0028125 &0.0022044&0.0023131 & \\ \hline
5f&0.025&0.0094531 &0.0091507&0.0091521 &  \\ \hline
    &0.050&0.0028125 &0.0017421&0.0017835 &  \\ \hline
5g&0.025&0.0094531 &0.0089465&0.0089465 & \\ \hline
    &0.050&0.0028125 &0.0010664&0.0010159 & \\ \hline
  6p &0.025&0.0042014&0.0041493&0.0041548&  \\ \hline
6d &0.025& 0.0042014&0.0040452&0.0040606& \\ \hline
 6f&0.025&0.0042014&0.0038901&0.0039168&\\ \hline
 6g&0.025& 0.0042014&0.0036943&0.0037201&
\\ \hline

\end{tabular}

\caption{Energy eigenvalues of the Hulthe\'n potential as a function
of the screening parameter for
2p, 3p, 3d, 4p, 4d, 4f, 5p, 5d, 5f, 5g, 6p, 6d, 6f and 6g states in atomic units
($\hbar=m=e=1$) and for $Z=1$.} \label{table2}
\end{table}

\newpage

\begin{thebibliography}{99}
\bibitem{Greiner}
W. Greiner, Quantum Mechanics, 4th. edn. (Springer, Berlin, 2001).
\bibitem{Bagrov}
V. G. Bagrov, D. M. Gitman, Exact Solutions of Relativistic Wave
Equations (Kluwer Academic Publishers, Dordrecht, 1990).
\bibitem{Eckart}
C.~Eckart, Phys. Rev. A\textbf{41}, 4682 (1930).
\bibitem{Hulten1}
 L. Hulth\'en , Ark. Mat. Astron. Fys. A\textbf{28}, 5 (1942).
\bibitem{Hulten2}
 L. Hulthe\'n, Ark. Mat. Astron. Fys.  B\textbf{29}, 1 (1942).
\bibitem{Flugge}
S. Fl\"{u}gge , Practical Quantum Mechanics, Vol. 1 (Springer,
Berlin, 1994).
\bibitem{Lam}
C. S. Lam, Y.P. Varshni, Phys. Rev. A\textbf{4}, 1875 (1971).
\bibitem{Varshni}
Y. P. Varshni, Phys. Rev. A\textbf{41}, 4682 (1990).
\bibitem{Lai1}
C. S. Lai, W.C. Lin, Phys. Lett. A\textbf{78}, 335 (1980).
\bibitem{Dutt}
 R. Dutt, U. Mukherji, Phys. Lett. A \textbf{90}, 395 (1982).
\bibitem{Patil}
S. H. Patil, J. Phys. A\textbf{17}, 575 (1984).
\bibitem{Popov}
 V. S. Popov, V. M. Weinberg, Phys. Lett. A\textbf{107}, 371 (1985).
\bibitem{Roy}
 B. Roy, R. Roychoudhury, J. Phys. A\textbf{20},  3051 (1987).
\bibitem{Tang}
 A. Z. Tang, F. T. Chan, Phys. Rev. A\textbf{35}, 911 (1987).
\bibitem{Lai2}
 C. H. Lai, J. Math. Phys. \textbf{28}, 1801 (1987).
\bibitem{Matthys}
 P. Matthys, H. De Meyer, Phys. Rev. A\textbf{38}, 1165 (1988).
\bibitem{Laha}
U. Laha, C. Bhattacharyya, K. Roy, B. Talukdar, Phys. Rev.
C\textbf{38}, 558 (1988).
\bibitem{Talukdar}
B. Talukdar, U. Das, C. Bhattacharyya, P. K. Bera, J. Phys. A
\textbf{25}, 4073 (1992).
\bibitem{Filho}
E. D. Filho, R. M. Ricotta, Mod. Phys. Lett. A\textbf{10}, 1613
(1995).
\bibitem{Gonul}
B. G\"{o}n\"{u}l, O. \"{O}zer, Y.Cancelik and M.Kocak, Phys. Lett.
A\textbf{275}, 238 (1995).
\bibitem{Nikiforov}
 A. F. Nikiforov and  V. B. Uvarov, Special Functions of Mathematical Physics, (Birkh\"{a}user, Basel 1988).
\bibitem{Gendenshtein}
L. Gendenshtein, JETP Lett. \textbf{38}, 356 (1983).
\bibitem{Badalov1}
V. H. Badalov, H. I. Ahmadov, and A. I. Ahmadov, Int. J. Mod. Phys.
E \textbf{18}, 631 (2009).
\bibitem{Badalov2}
V. H. Badalov, H. I. Ahmadov, and S. V. Badalov, Int. J. Mod. Phys.
E \textbf{19}, 1463  (2010).
\bibitem{Badalov3}
V. H. Badalov, H. I. Ahmadov, math-ph/1111.4734.
\bibitem{Ahmadov1}
H. I. Ahmadov, C. Aydin, N. Sh. Huseynova, and O. Uzun. Int. J. Mod.
Phys. E \textbf{22}, 1350072 (2013).
\bibitem{Ahmadov2}
 A. I. Ahmadov, C. Aydin and O. Uzun. Int. J. Modern. Phys. A\textbf{29},
1450002  (2014).
\bibitem{Feizi}
H.~Feizi, M. R.~Shojaei and A. A.~Rajabi, Adv. Studies Theor. Phys.
\textbf{6}, 477 (2012).
\bibitem{Jia}
C. S. Jia, T. Chen and L. G. Cui, Phys. Lett. A \textbf{373},
 (2009) 1621.
\bibitem{Greene}
R. L. Greene and C. Aldrich, Phys. Rev. A \textbf{14} (1976) 2363.
\bibitem{Area}
I. Area, E. Godoy, A. Ronveaux, A. Zarzo, J.  Comput. Appl. Math.
\textbf{157} (2003) 93.
\bibitem{Ikhdair}
S. M. ~Ikhdair and R. Sever, J. of. Math. Chem. \textbf{42}, 461
(2007).
\bibitem{Abramowitz}
M. Abramowitz, I. A. Stegun, Handbook of Mathematical Functions with
Formulas, Graphs and Mathematical Tables (Dover, New York, 1964).
\bibitem{Agboola}
D. Agboola, Commun. Theor. Phys. \textbf{55} (2011) 972.
\bibitem{Cooper1}
F. Cooper, A.~Khare, U.~Sukhatme, Supersymmetry in Quantum Mechnics,
World Scientific, 2001.
\bibitem{Cooper2}
F. Cooper, A.~Khare, U.~Sukhatme, Phys. Rep. {\bf 251}, 367 (1995).
\bibitem{Boztosun}
O. Bayrak, G. Kocak and I. Boztosun, J. Phys. A: Math. Gen.
\textbf{39}, 11521 (2006).

\end{thebibliography}
\end{document}